\documentstyle[fleqn,11pt]{book}
\begin{document}
\parindent 1.4cm
\large
\begin{center}
{\Large \bf THE ERMAKOV-LEWIS INVARIANTS OF THE SCHR\"{O}DINGER
EQUATION CONTINUOUS MEASUREMENT}
\end{center}
\begin{center}
{{\bf Jos\'e Maria Filardo Bassalo}}
\end{center}
\begin{center}
{Funda\c{c}\~ao Minerva}
\end{center}
\begin{center}
{{\bf Paulo de Tarso Santos Alencar}}
\end{center}
\begin{center}
{Professor Aposentado da UFPA}
\end{center}
\begin{center}
{{\bf Daniel Gemaque da Silva}}
\end{center}
\begin{center}
{Professor de Ensino M\'edio, Amap\'a}
\end{center}
\begin{center}
{{\bf Antonio Boulhosa Nassar}}
\end{center}
\begin{center}
{Extension Program-Department of Sciences, University of California,\
Los Angeles, California 90024}
\end{center}
\begin{center}
{{\bf M. Cattani}}
\end{center}
\begin{center}
{Instituto de F\'{\i}sica da USP,\ 05389-970,\ S\~ao Paulo,\ SP}
\end{center}
\par
ABSTRACT:\ In this work we study the Ermakov-Lewis invariants of
the Schr\"{o}dinger equation continuous measurement.
\vspace{0.2cm}
\par
PACS 03.65\ -\ Quantum Mechanics
\vspace{0.2cm}
\par
1.\ {\bf Introduction}
\par
Many years ago, in 1967[1], H. R. Lewis has shown that there is a
conserved quantity, that will be indicated by $I$, associated with the
time dependent harmonic oscillator ($TDHO$) with frequency ${\omega}(t)$,
given by:
\begin{center}
{$I\ =\ {\frac {1}{2}}[({\dot {q}}\ {\alpha}\ -\ {\dot {{\alpha}}}\
q)^{2}\ +\ ({\frac {q}{{\alpha}}})^{2}]$\ ,\ \ \ \ \ (1.1)}
\end{center}
where $q$ and ${\alpha}$ obey, respectively the equations:
\begin{center}
{${\ddot {q}}\ +\ {\omega}^{2}(t)\ q\ =\ 0\ ,\ \ \ \ \ {\ddot
{{\alpha}}}\ +\ {\omega}^{2}(t)\ {\alpha}\ =\ {\frac
{1}{{\alpha}^{3}}}$\ .\ \ \ \ \ (1.2,3)}
\end{center}
\par
On the other hand, as the above expressions have also been obtained by
V. P. Ermakov [2] in 1880, the invariants determination of time
dependent physical systems is also known as the {\bf Ermakov-Lewis
problem}. So, considerable efforts have been devoted to solve this
problem and its generalizations, in the last thirty years, and in many
works have been published on these subjects [3,4].
\par
In the present work we investigate the existence of these invariants
for the one-dimensional Schr\"{o}dinger equation describing one system
undergoing continuous measurement.
\par
2.\ {\bf Schr\"{o}dinger Equation Continuous Measurement}
\par
The Schr\"{o}dinger equation describing one system undergoing
continuous measurement, is given by [5]:
\begin{center}
{i\ ${\hbar}\ {\frac {{\partial}{\psi}(x,\ t)}{{\partial}t}}\ =\ -\
{\frac {{\hbar}^{2}}{2\ m}}\ {\frac {{\partial}^{2}\ {\psi}(x,\
t)}{{\partial}x^{2}}}\ +\ {\Big {[}}\ {\frac {1}{2}}\ m\ {\omega}^{2}\
x^{2}\ +\ {\lambda}\ x\ X(t)\ {\Big {]}}\ {\psi}(x,\ t)\ -$}
\end{center}
\begin{center}
{$-\ {\frac {i\ {\hbar}}{4\ {\tau}}}\ {\Bigg {(}}\ {\frac {[x\ -\ {\bar
{x}}(t)]^{2}}{{\delta}^{2}(t)}}\ -\ 1\ {\Bigg {)}}\ {\psi}(x,\
t)$\ ,\ \ \ \ \ (2.1)}
\end{center}
where ${\psi}(x,\ t)$ is a wavefunction which describes the
system considered, $X(t)$ is a classical position of one particle
in a harmonic potential of frequency ${\omega}$, ${\tau}$ and
${\delta}$ have dimensions of time and space, respectively, and
${\bar {x}}(t)$ means a mean value $<\ x(t)\ >$.
\par
Writing the wavefuncition ${\psi}(x,\ t)$ in the polar form, defined by
the Madelung-Bohm transformation[6,7], we get:
\begin{center}
{${\psi}(x,\ t)\ =\ {\phi}(x,\ t)\ e^{i\ S(x,\ t)}$\ ,\ \ \ \ \ (2.2)}
\end{center}
where $S(x\ ,t)$ is the classical action and ${\phi}(x,\ t)$ will be
defined in what follows.
\par
Substituting Eq.(2.2) into Eq.(2.1) and taking the real and imaginary
parts of the resulting equation, we get[8]:
\par
\begin{center}
{${\frac {{\partial}{\rho}}{{\partial}t}}\ +\ {\frac
{{\partial}({\rho}\ v_{qu})}{{\partial}x}}\ =\ -\ {\frac {{\rho}}{2\
{\tau}}}\ {\Bigg {(}}\ {\frac {[x\ -\ {\bar
{x}}(t)]^{2}}{{\delta}^{2}(t)}}\ -\ 1\ {\Bigg {)}}$\ ,\ \ \ \ \ (2.3)}
\end{center}
\begin{center}
{${\frac {{\partial}v_{qu}}{{\partial}t}}\ +\ v_{qu}\ {\frac
{{\partial}v_{qu}}{{\partial}x}}\ +\ {\omega}^{2}\ x\ +\ {\frac
{{\lambda}}{m}}\ X(t)\ =\ -\ {\frac {1}{m}}\ {\frac
{{\partial}\ V_{qu}}{{\partial}x}}$\ ,\ \ \ \ \ (2.4)}
\end{center}
where:
\begin{center}
{${\rho}(x,\ t)\ =\ {\phi}^{2}(x,\ t)$\ ,\ \ \ \ \ (2.5)\ \ \
(quantum mass density)}
\end{center}
\begin{center}
{$v_{qu}(x,\ t)\ =\ {\frac {{\hbar}}{m}}\ {\frac {{\partial}S(x,\
t)}{{\partial}x}}$\ ,\ \ \ \ \ (2.6)\ \ \ \ \ (quantum velocity)}
\end{center}
and
\begin{center}
{$V_{qu}(x,\ t)\ =\ -\ {\frac {{\hbar}^{2}}{2\ m}}\ {\frac {1}{{\sqrt
{{\rho}}}}}\ {\frac {{\partial}^{2}{\sqrt
{{\rho}}}}{{\partial}x^{2}}}$\ .\ \ \ \ \ (2.7)\ \ \ \ \ (Bohm quantum
potential)}
\end{center}
\par
In order to integrate Eq.(2.4) let us assume that the expected value of
{\bf quantum force} is equal to zero for all times $t$, that is,
\begin{center}
{$<\ {\frac {{\partial}V_{qu}}{{\partial}x}}\ >\ \ \ {\to}\ \  0\ \ \
{\Leftrightarrow}\ \ \ {\frac {{\partial}V_{qu}}{{\partial}x}}\
{\vert}_{x\ =\ {\bar {x}}(t)}\ ,\ \ \ \ \ <\ x\ >\ =\ {\bar {x}}(t)$\
.\ \ \ \ \ \ (2.9a-c)}
\end{center}
\par
In this way, we can write Eq.(2.4) into two parts:
\begin{center}
{${\frac {{\partial}v_{qu}}{{\partial}t}}\ +\ v_{qu}\ {\frac
{{\partial}v_{qu}}{{\partial}x}}\ +\ {\omega}^{2}\ x\ +\ {\frac
{{\lambda}}{m}}\ X(t)\ =\ k(t)\ [x\ -\ {\bar {x}}(t)]$\ ,\ \ \ \ \ (2.10)}
\end{center}
\begin{center}
{$-\ {\frac {1}{m}}\ {\frac {{\partial}\ V_{qu}}{{\partial}x}}\ =\
{\frac {{\partial}}{{\partial}x}}\ {\Big {(}}\ {\frac {{\hbar}^{2}}{2\
m^{2}}}\ {\frac {1}{{\sqrt {{\rho}}}}}\ {\frac {{\partial}^{2}{\sqrt
{{\rho}}}}{{\partial}x^{2}}}\ {\Big {)}}\ =\ k(t)\ [x\ -\ {\bar
{x}}(t)]$\ .\ \ \ \ \ (2.11)}
\end{center}
\par
Performing the differentiations indicated in Eq.(2.11) we get,
\begin{center}
{${\frac {{\hbar}^{2}}{4\ m^{2}}}\ [{\frac {1}{{\rho}}}\ {\frac
{{\partial}^{3}{\rho}}{{\partial}x^{3}}}\ -\ {\frac {2}{{\rho}^{2}}}\
{\frac {{\partial}{\rho}}{{\partial}x}}\ {\frac
{{\partial}^{2}{\rho}}{{\partial}x^{2}}}\ +\ {\frac {1}{{\rho}^{3}}}\
({\frac {{\partial}{\rho}}{{\partial}x}})^{3}]\ =\ k(t)\ [x\ -\ {\bar
{x}}(t)]$\ .\ \ \ \ \ (2.12)}
\end{center}
\par
To integrate Eq.(2.12) it is necessary to known the initial condition
for ${\rho}(x,\ t)$. Let us assume that for $t\ =\ 0$ the physical
system is represented by a normalized Gaussian wave packet, centered at
${\bar {x}}(0)$, that is,
\begin{center}
{${\rho}(x,\ 0)\ =\ [2\ {\pi}\ {\delta}^{2}(0)]^{-\ 1/2}\ e^{-\ {\frac
{[x\ -\ {\bar {x}}(0)]^{2}}{2\ {\delta}^{2}(0)}}}\ =\ {\frac {1}{{\sqrt
{A}}}}\ e^{-\ {\frac {B^{2}}{C}}}$\ ,\ \ \ \ \ (2.13)}
\end{center}
\begin{center}
{$A\ =\ 2\ {\pi}\ {\delta}^{2}(0)\ ,\ \ \ \ \ B\ =\ x\ -\ {\bar {x}}(0)\ ,\ \
\ \ \ C\ =\ 2\ {\delta}^{2}(0)$\ .\ \ \ \ \ (2.14-16)}
\end{center}
\par
Since Eq.(2.13) is a particular solution of Eq.(2.12), we must have:
\begin{center}
{${\frac {{\hbar}^{2}}{4\ m^{2}}}\ {\Big {[}}\ {\frac {1}{{\rho}(x,\
0)}}\ {\frac {{\partial}^{3}{\rho}(x,\ 0)}{{\partial}x^{3}}}\ -\ {\frac
{2}{{\rho}^{2}(x,\ 0)}}\ {\frac {{\partial}{\rho}(x,\
0)}{{\partial}x}}\ {\frac {{\partial}^{2}{\rho}(x,\
0)}{{\partial}x^{2}}}\ +\ {\frac {1}{{\rho}(x,\ 0)^{3}}}\ {\Big {(}}\ {\frac
{{\partial}{\rho}(x,\ 0)}{{\partial}x}}\ {\Big {)}}^{3}\ {\Big {]}}\ =$}
\end{center}
\begin{center}
{$=\ k(0)\ [x\ -\ {\bar {x}}(0)]$\ .\ \ \ \ \ (2.17)}
\end{center}
\par
Performing the differentiation indicated above, Eq.(2.17) becomes:
\begin{center}
{${\frac {{\hbar}^{2}\ [x\ -\ {\bar {x}}(0)]}{4\ m^{2}\ {\delta}^{4}(0)}}\
\ =\ k(0)\ [x\ - {\bar {x}}(0)]\ \ \ {\to}\ \ \ k(0)\ =\ {\frac
{{\hbar}^{2}}{4\ m^{2}\ {\delta}^{4}(0)}}$\ .\ \ \ \ \ (2.18)}
\end{center}
\par
Comparing Eq.(2.18) with the Eqs.(2.12) and (2.13), by analogy we get,
\begin{center}
{$k(t)\ =\ {\frac {{\hbar}^{2}}{4\ m^{2}\ {\delta}^{4}(t)}}\ \ \ {\to}\ \
\ {\delta}^{4}(t)\ =\ {\frac {{\hbar}^{2}}{4\ m^{2}\ k(t)}}$\ ,\ \ \ \ \
(2.19,20)}
\end{center}
\begin{center}
{${\rho}(x,\ t)\ =\ [2\ {\pi}\ {\delta}^{2}(t)]^{-\ 1/2}\ e^{-\ {\frac {[x\ -\
{\bar {x}}(t)]^{2}}{2\ {\delta}^{2}(t)}}}$\ ,\ \ \ \ \ \ (2.21)}
\end{center}
\par
Taking into account Eqs.(2.19-21), let us perform the following
differentiations, remembering that $t$ and $x$ are independent variables:
\begin{center}
{${\frac {{\partial}{\rho}}{{\partial}t}}\ =\ {\Big {[}}\ -\ {\frac
{{\dot {{\delta}}}}{{\delta}}}\ +\ {\frac {(x\ -\ {\bar
{x}})}{{\delta}^{2}}}\ {\dot {{\bar {x}}}}\ +\ {\frac {2\ (x\ -\
{\bar {x}})^{2}}{{\delta}^{3}}}\ {\rho}\ {\Big {]}}$\ ,\ \ \ \ \
(2.22)}
\end{center}
\begin{center}
{${\frac {{\partial}{\rho}}{{\partial}x}}\ =\ -\ {\frac {(x\ -\
{\bar {x}})}{{\delta}^{2}}}\ {\rho}$\ .\ \ \ \ \ (2.23)}
\end{center}
\par
>From Eq.(2.3) and Eqs.(2.22-23), results
\begin{center}
{${\frac {{\partial}{\rho}}{{\partial}t}}\ +\ {\frac
{{\partial}({\rho}\ v_{qu})}{{\partial}x}}\ =\ -\ {\frac {{\rho}}{2\
{\tau}}}\ {\Bigg {(}}\ {\frac {[x\ -\ {\bar
{x}}(t)]^{2}}{{\delta}^{2}(t)}}\ -\ 1\ {\Bigg {)}}\ =$}
\end{center}
\begin{center}
{$=\ {\frac {{\partial}v_{qu}}{{\partial}x}}\ -\ {\frac
{v_{qu}}{{\delta}^{2}}}\ (x\ -\ {\bar {x}})\ =\ {\frac {{\dot
{{\delta}}}}{{\delta}}}\ -\ {\frac {{\dot
{{\delta}}}}{{\delta}^{3}}}\ (x\ -\ {\bar {x}})^{2}\ -\ {\frac {{\dot
{{\bar {x}}}}}{{\delta}^{2}}}\ (x\ -\ {\bar {x}})\ -$}
\end{center}
\begin{center}
{$-\ {\frac {1}{2\ {\tau}\ {\delta}^{2}}}\ (x\ -\ {\bar {x}})^{2}\ +\
{\frac {1}{2\ {\tau}}}$\ .\ \ \ \ \ (2.24)}
\end{center}
\par
Defining
\begin{center}
{$p(x,\ t)\ =\ -\ {\frac {(x\ -\ {\bar {x}})}{{\delta}^{2}}}$\ ,\ \ \ \ \
(2.25)}
\end{center}
\begin{center}
{$r(x,\ t)\ =\ {\frac {{\dot {{\delta}}}}{{\delta}}}\ -\ {\frac
{{\dot {{\delta}}}}{{\delta}^{3}}}\ (x\ -\ {\bar {x}})^{2}\ -\ {\frac
{(x\ -\ {\bar {x}})}{{\delta}^{2}}}\ {\dot {{\bar {x}}}}\ -\ {\frac
{1}{2\ {\tau}\ {\delta}^{2}}}\ (x\ -\ {\bar {x}})^{2}\ +\ {\frac
{1}{2\ {\tau}}}$\ ,\ \ \ \ \ (2.26)}
\end{center}
Eq.(2.24) becomes,
\begin{center}
{${\frac {{\partial}v_{qu}}{{\partial}x}}\ +\ p(x,\ t)\ v_{qu}\ =\
r(x,\ t)$\ ,\ \ \ \ \ (2.27)}
\end{center}
which can be integrated, giving:
\begin{center}
{$v_{qu}\ =\ {\frac {1}{u}}\ [{\int}\ r\ u\ {\partial}x\ +\ c(t)]$\ ,\
\ \ \ \ (2.28)}
\end{center}
where,
\begin{center}
{$u\ =\ exp\ ({\int}\ p\ {\partial}x)$\ .\ \ \ \ \ (2.29)}
\end{center}
\par
Using Eqs.(2.20,28,29), the function $u$ given by Eq.(2.29) is written
us[9]:
\begin{center}
{$u\ =\ exp\ {\Big {(}}\ {\int}\ [-\ {\frac {(x\ -\ {\bar
{x}})}{{\delta}^{2}}}]\ {\partial}x\ {\Big {)}}\ =\ ({\pi}\
{\delta}^{2})^{1/2}\ {\rho}$\ .\ \ \ \ \ (2.30)}
\end{center}
\par
In this way, defining $I\ =\ {\int}\ r\ u\ {\partial}x$, and using
Eqs.(2.26,30) we obtain:
\begin{center}
{$I\ =\ {\int}\ r\ u\ {\partial}x\ =$}
\end{center}
\begin{center}
{$=\ {\int}\ {\Big {[}}\ {\frac {{\dot {{\delta}}}}{{\delta}}}\ -\ {\frac
{{\dot {{\delta}}}}{{\delta}^{3}}}\ (x\ -\ {\bar {x}})^{2}\ -\ {\frac
{(x\ -\ {\bar {x}})}{{\delta}^{2}}}\ {\dot {{\bar {x}}}}\ -\ {\frac
{1}{2\ {\tau}\ {\delta}^{2}}}\ (x\ -\ {\bar {x}})^{2}\ +\ {\frac
{1}{2\ {\tau}}}\ {\Big {]}}\ ({\pi}\ {\delta}^{2})^{1/2}\ {\rho}\
{\partial}x\ \ \ {\to}$}
\end{center}
\begin{center}
{$I\ =\ I_{1}\ +\ I_{2}\ +\ I_{3}$\ ,\ \ \ \ \ (2.31)}
\end{center}
where,
\begin{center}
{$I_{1}\ =\ {\int}\ {\Big {[}}\ {\frac {{\dot {{\delta}}}}{{\delta}}}\ -\
{\frac {{\dot {{\delta}}}}{{\delta}^{3}}}\ (x\ -\ {\bar {x}})^{2}\
{\Big {]}}\ ({\pi}\ {\delta}^{2})^{1/2}\ {\rho}\ {\partial}x$\ ,\ \ \ \
\ (2.32)}
\end{center}
\begin{center}
{$I_{2}\ =\ {\int}\ {\Big {[}} -\ {\frac {(x\ -\ {\bar
{x}})}{{\delta}^{2}}}\ {\dot {{\bar {x}}}}\ {\Big {]}}\ ({\pi}\
{\delta}^{2})^{1/2}\ {\rho}\ {\partial}x$\ ,\ \ \ \ \ (2.33)}
\end{center}
and
\begin{center}
{$I_{3}\ =\ {\int}\ {\Big {[}}\ -\ {\frac
{1}{2\ {\tau}\ {\delta}^{2}}}\ (x\ -\ {\bar {x}})^{2}\ +\ {\frac
{1}{2\ {\tau}}}\ {\Big {]}}\ ({\pi}\
{\delta}^{2})^{1/2}\ {\rho}\ {\partial}x$\ .\ \ \ \ \ (2.34)}
\end{center}
\par
To integrate Eq.(2.32) it is necessary, first to perform the
differentiation[9] shown bellow, where Eq.(2.23) is used:
\begin{center}
{${\frac {{\partial}}{{\partial}x}}\ [{\frac {{\dot
{{\delta}}}}{{\delta}}}\ (x-\ {\bar {x}})\ {\rho}]\ =\ [{\frac {{\dot
{{\delta}}}}{{\delta}}}\ -\ {\frac {{\dot {{\delta}}}\ (x\ -\ {\bar
{x}})^{2}}{{\delta}^{3}}}]\ {\rho}$\ .\ \ \ \ \ (2.35)}
\end{center}
\par
Inserting Eq.(2.35) into Eq.(2.32), results
\begin{center}
{$I_{1}\ =\ ({\pi}\ {\delta}^{2})^{1/2}\ {\rho}\ ({\frac {{\dot
{{\delta}}}}{{\delta}}})\ (x\ -\ {\bar {x}})$\ .\ \ \ \ \ (2.36)}
\end{center}
\par
Similarly, to calculate $I_{2}$, seen in Eq.(2.33), we need to use
Eq.(2.23) obtaining,
\begin{center}
{$I_{2}\ =\ ({\pi}\ {\delta}^{2})^{1/2}\ {\dot {{\bar {x}}}}\ {\rho}$\
.\ \ \ \ \ (2.37)}
\end{center}
\par
Now, inserting Eq. (2.21) into Eq. (2.34), and remember that ${\int}\
u^{2}\ exp\ (-\ u^{2})\ du\ =\ {\sqrt {{\pi}}}/2$ and ${\int}\ exp\ (-\
u^{2})\ du\ =\ {\sqrt {{\pi}}}/2$, results
\begin{center}
{$I_{3}\ =\ 0$\ .\ \ \ \ \ \ (2.38)}
\end{center}
\par
Substituting Eqs.(2.36-38) into Eq.(2.31) we see that,
\begin{center}
{$I\ =\ ({\pi}\ {\delta}^{2})^{1/2}\ {\rho}\ [{\frac {{\dot
{{\delta}}}}{{\delta}}}\ (x\ -\ {\bar {x}})\ +\ {\dot {{\bar {x}}}}]$\
.\ \ \ \ \ (2.39)}
\end{center}
\par
Remembering that the quantum velocity $v_{qu}$ is defined by Eq.(2.28)
and using Eqs.(2.30,39) we verify that $v_{qu}$ can be written as:
\begin{center}
{$v_{qu}\ =\ {\frac {{\dot {{\delta}}}}{{\delta}}}\ (x\ -\ {\bar {x}})\
+\ {\dot {{\bar {x}}}}\ +\ {\frac {c(t)}{({\pi}\ {\delta}^{2})^{1/2}\
{\rho}}}$\ .\ \ \ \ \ (2.40)}
\end{center}
\par
Assuming that the mass density ${\rho}\ {\to}\ 0$ when ${\mid}\ x\ {\mid}\
{\to}\ {\infty}$ we verify that the parameter $c(t)$ must be equal to
zero. Consequently, $v_{qu}$ becomes,
\begin{center}
{$v_{qu}(x,\ t)\ =\ {\frac {{\dot {{\delta}}}}{{\delta}}}\ (x\ -\ {\bar
{x}})\ +\ {\dot {{\bar {x}}}}$\ .\ \ \ \ \ (2.41)}
\end{center}
\par
However, as Eq. (2.41) is not compatible with Eq. (2.3), we must
modified Eq. (2.41) for:
\begin{center}
{$v_{qu}(x,\ t)\ =\ {\Big {(}}\ {\frac {{\dot {{\delta}}}}{{\delta}}}\
+\ {\frac {1}{2\ {\tau}}}\ {\Big {)}}\ (x\ -\ {\bar {x}})\ +\ {\dot
{{\bar {x}}}}$\ ,\ \ \ \ \ (2.42)}
\end{center}
which satisfies Eq. (2.3).
\par
Now, using the above Eq.(2.42) we calculate the following differentiations,
remembering that $t$ and $x$ as independent variables:
\begin{center}
{${\frac {{\partial}v_{qu}}{{\partial}t}}\ =\ {\frac
{{\partial}}{{\partial}t}}\ {\Big {[}}\ {\Big {(}}\ {\frac {{\dot
{{\delta}}}}{{\delta}}}\ +\ {\frac {1}{2\ {\tau}}}\ {\Big {)}}\ (x\ -\
{\bar {x}})\ +\ {\dot {{\bar {x}}}}\ {\Big {]}}\ =$}
\end{center}
\begin{center}
{$=\ {\Big {(}}\ {\frac {{\ddot {{\delta}}}}{{\delta}}}\ -\ {\frac
{({\dot {{\delta}}})^{2}}{{\delta}^{2}}}\ {\Big {)}}\ (x\ -\ {\bar {x}})\ -\
{\Big {(}}\ {\frac {{\dot {{\delta}}}}{{\delta}}}\ +\ {\frac {1}{2\
{\tau}}}\ {\Big {)}}\ {\dot {{\bar {x}}}}\ +\ {\ddot {{\bar {x}}}}$\
,\ \ \ \ \ (2.43)}
\end{center}
\begin{center}
{${\frac {{\partial}v_{qu}}{{\partial}x}}\ =\ {\frac
{{\partial}}{{\partial}x}}\ {\Big {[}}\ {\Big {(}}\ {\frac {{\dot
{{\delta}}}}{{\delta}}}\ +\ {\frac {1}{2\ {\tau}}}\ {\Big {)}}\ (x\ -\
{\bar {x}})\ +\ {\dot {{\bar {x}}}}\ {\Big {]}}\ =\ {\frac {{\dot
{{\delta}}}}{{\delta}}}\ +\ {\frac {1}{2\ {\tau}}}$\ .\ \ \ \ \ (2.44)}
\end{center}
\par
Now, using Eqs. (2.19,42-44) in Eq. (2.10) and adding the factor
${\omega}^{2}\ {\bar {x}}$, we have
\begin{center}
{${\Big {[}} {\frac {{\ddot {{\delta}}}}{{\delta}}}\ -\ {\frac {({\dot
{{\delta}}})^{2}}{{\delta}^{2}}}\ {\Big {]}}\ (x\ -\ {\bar {x}})\ -\
{\Big {(}}\ {\frac {{\dot {{\delta}}}}{{\sigma}}}\ +\ {\frac {1}{2\
{\tau}}}\ {\Big {)}}\ {\dot {{\bar {x}}}}\ +\ {\ddot {{\bar {x}}}}\
+\ {\Big {[}}\ {\Big {(}}\ {\frac {{\dot {{\delta}}}}{{\delta}}}\ +\
{\frac {1}{2\ {\tau}}}\ {\Big {)}}\ (x\ -\ {\bar {x}})\ +\ {\dot {{\bar
{x}}}}\ {\Big {]}}\ {\Big {(}}\ {\frac {{\dot {{\delta}}}}{{\delta}}}\
+\ {\frac {1}{2\ {\tau}}}\ {\Big {)}}\ +$}
\end{center}
\begin{center}
{$+\ {\omega}^{2}\ x\ +\ {\omega}^{2}\ {\bar {x}}\ -\ {\omega}^{2}\
{\bar {x}}\ +\ {\frac {{\lambda}}{m}}\ X(t)\ =\ {\frac
{{\hbar}^{2}}{4\ m^{2}\ {\delta}^{4}}}\ (x\ - {\bar {x}})\ \ \ {\to}$}
\end{center}
\begin{center}
{${\Big {(}}\ {\frac {{\ddot {{\delta}}}}{{\delta}}}\ +\ {\frac {{\dot
{{\delta}}}}{{\delta}\ {\tau}}}\ +\ {\frac {1}{4\ {\tau}^{4}}}\ +\
{\omega}^{2}\ -\ {\frac {{\hbar}^{2}}{4\ m^{2}\ {\delta}^{4}}}\ {\Big
{)}}\ (x\ -\ {\bar {x}})\ +\ {\ddot {{\bar {x}}}}\ +\ {\omega}^{2}\
{\bar {x}}\ +\ {\frac {{\lambda}}{m}}\ X(t)\ =\ 0$\ .\ \ \ \ \
(2.45)}
\end{center}
\par
To satisfy Eq.(2.45), the following conditions must be obeyed:
\begin{center}
{${\ddot {{\delta}}}\ +\ {\frac {{\dot {{\delta}}}}{{\tau}}}\ +\ {\Big
{(}}\ {\omega}^{2}\ +\ {\frac {1}{4\ {\tau}^{4}}}\ {\Big {)}}\ {\delta}\ =\
{\frac {{\hbar}^{2}}{4\ m^{2}\ {\delta}^{3}}}$\ ,\ \ \ \ \ (2.46)}
\end{center}
\begin{center}
{${\ddot {{\bar {x}}}}\ +\ {\omega}^{2}\ {\bar {x}}\ +\ {\frac
{{\lambda}}{m}}\ X(t)\ =\ 0$\ .\ \ \ \ \ (2.47)}
\end{center}
\par
Putting
\begin{center}
{${\delta}(t)\ =\ {\Big {(}}\ {\frac {{\hbar}^{2}}{4\ m^{2}}}\ {\Big
{)}}^{1/4}\ {\alpha}(t)$\ ,\ \ \ \ \ (2.48)}
\end{center}
in Eq. (2.46), we obtain,
\begin{center}
{${\ddot {{\alpha}}}\ +\ {\frac {{\dot {{\alpha}}}}{{\tau}}}\ +\
{\Big {(}} {\omega}^{2}\ +\ {\frac {1}{4\ {\tau}^{4}}}\ {\Big
{)}}\ {\alpha}\ =\ {\frac {1}{{\alpha}^{3}}}$\ .\ \ \ \ \ (2.49)}
\end{center}
\par
Finally, eliminating the factor ${\omega}^{2}$ into Eqs.(2.47) and
(2.49) we get,
\begin{center}
{${\ddot {{\alpha}}}\ +\ {\frac {{\dot {{\alpha}}}}{{\tau}}}\ +\
{\Big {(}}\ -\ {\frac {{\ddot {{\bar {x}}}}\ +\ {\lambda}\
X(t)/m}{{\bar {x}}}}\ +\ {\frac {1}{4\ {\tau}^{4}}}\ {\Big {)}}\
{\alpha}\ =\ {\frac {1}{{\alpha}^{3}}}\ \ \ {\to}$}
\end{center}
\begin{center}
{${\bar {x}}\ {\ddot {{\alpha}}}\ -\ {\ddot {{\bar {x}}}}\ {\alpha}\ +\
{\Big {(}}\ {\frac {{\dot {{\alpha}}}}{{\alpha}\ {\tau}}}\ -\ {\frac
{{\lambda}\ X(t)}{m\ {\bar {x}}}}\ +\ {\frac {1}{4\ {\tau}^{4}}}\ {\Big
{)}}\ {\alpha}\ {\bar {x}}\ =\ {\frac {{\bar {x}}}{{\alpha}^{3}}}\ \ \
{\to}$}
\end{center}
\begin{center}
{${\frac {d}{dt}}\ ({\dot {{\alpha}}}\ {\bar {x}}\ -\ {\dot {{\bar
{x}}}}\ {\alpha})\ =\ {\frac {{\bar {x}}}{{\alpha}^{3}}}\ -\ {\Big {(}}\
{\frac {{\dot {{\alpha}}}}{{\alpha}\ {\tau}}}\ -\ {\frac {{\lambda}\ X(t)}{m\
{\bar {x}}}}\ +\ {\frac {1}{4\ {\tau}^{4}}}\ {\Big {)}}\ {\alpha}\
{\bar {x}}\ \ \ \ {\to}$}
\end{center}
\begin{center}
{$({\dot {{\alpha}}}\ {\bar {x}}\ -\ {\dot {{\bar {x}}}}\ {\alpha})\
{\frac {d}{dt}}\ ({\dot {{\alpha}}}\ {\bar {x}}\ -\ {\dot {{\bar
{x}}}}\ {\alpha})\ -\ {\frac {{\bar {x}}}{{\alpha}^{3}}}\ ({\dot
{{\alpha}}}\ {\bar {x}}\ -\ {\dot {{\bar {x}}}}\ {\alpha})\ =$}
\end{center}
\begin{center}
{$=\ -\ {\Big {(}}\ {\frac {{\dot {{\alpha}}}}{{\alpha}\ {\tau}}}\ -\
{\frac {{\lambda}\ X(t)}{m\ {\bar {x}}}}\ +\ {\frac {1}{4\
{\tau}^{4}}}\ {\Big {)}}\ {\alpha}\ {\bar {x}}\ ({\dot {{\alpha}}}\
{\bar {x}}\ -\ {\dot {{\bar {x}}}}\ {\alpha})\ \ \ {\to}$}
\end{center}
\begin{center}
{${\frac {d}{dt}}\ {\Bigg {(}}\ {\frac {1}{2}}\ {\Big {[}}\ {\Big {(}}\
{\dot {{\alpha}}}\ {\bar {x}}\ -\ {\dot {{\bar {x}}}}\ {\alpha}\ {\Big
{)}}^{2}\ +\ {\Big {(}}\ {\frac {{\bar {x}}}{{\alpha}}}\ {\Big
{)}}^{2}\ {\Big {]}}\ {\Bigg {)}}\ =\ {\Big {(}}\ {\frac {{\dot
{{\alpha}}}}{{\alpha}\ {\tau}}}\ -\ {\frac {{\lambda}\ X(t)}{m\ {\bar
{x}}}}\ +\ {\frac {1}{4\ {\tau}^{4}}}\ {\Big {)}}\ {\alpha}^{3}\ {\bar
{x}}\ {\frac {d}{dt}}\ {\Big {(}}\ {\frac {{\bar {x}}}{{\alpha}}}\
{\Big {)}}\ \ \ {\to}$}
\end{center}
\begin{center}
{${\frac {dI}{dt}}\ =\ {\Big {(}}\ {\frac {{\dot
{{\alpha}}}}{{\alpha}\ {\tau}}}\ -\ {\frac {{\lambda}\ X(t)}{m\ {\bar
{x}}}}\ +\ {\frac {1}{4\ {\tau}^{4}}}\ {\Big {)}}\ {\alpha}^{3}\ {\bar
{x}}\ {\frac {d}{dt}}\ {\Big {(}}\ {\frac {{\bar {x}}}{{\alpha}}}\
{\Big {)}}$\ ,\ \ \ \ \ (2.50)}
\end{center}
where,
\begin{center}
{$I\ =\ {\frac {1}{2}}\ {\Big {[}}\ {\Big {(}}\ {\dot {{\alpha}}}\
{\bar {x}}\ -\ {\dot {{\bar {x}}}}\ {\alpha}\ {\Big {)}}^{2}\ +\ {\Big
{(}}\ {\frac {{\bar {x}}}{{\alpha}}}\ {\Big {)}}^{2}\ {\Big {]}}$\ ,\ \ \ \ \
(2.51)}
\end{center}
which represents the {\bf Ermakov-Lewis-Schr\"{o}dinger
invariant} of the time dependent harmonic oscillator ($TDHO$)[8].
In conclusion, we have shown that the {\bf Schr\"{o}dinger
equation continuos measurement} {\underline {does not have}} an
{\bf Ermakov-Lewis invariant} for the $TDHO$.[9]
\begin{center}
{{\bf NOTES AND REFERENCES}}
\end{center}
\par
1.\ LEWIS, H. R. 1967. {\it Physical Review Letters 18}, 510; 636 (E).
\par
2.\ ERMAKOV, V. P. 1880. {\it Univ. Izv. Kiev 20}, 1.
\par
3.\ NASSAR, A. B. 1986. {\it Journal of Mathematical Physics 27}, 755;
2949, and references therein.
4.\ ------ 1986. {\it Physical Review A33}, 2134, and references
therein.
\par
5.\ NASSAR, A. B. 2004. {\bf Chaotic Behavior of a Wave Packet under
Continuous Quantum Measurement} (mimeo).
\par
6.\ MADELUNG, E. 1926. {\it Zeitschrift f\"{u}r Physik 40}, 322.
\par
7.\ BOHM, D. 1952. {\it Physical Review 85}, 166.
\par
8.\ BASSALO, J. M. F., ALENCAR, P. T. S., CATTANI, M. S. D. e
NASSAR, A. B. 2003. {\bf T\'opicos da Mec\^anica Qu\^antica de de
Broglie-Bohm}, EDUFPA.
\par
9.\ In Eq. (2.50), we observe that if
\begin{center}
{$X(t)\ =\ {\frac {m}{{\lambda}}}\ {\Big {(}}\ {\frac {{\dot
{{\alpha}}}}{{\alpha}\ {\tau}}}\ +\ {\frac {1}{4\ {\tau}^{4}}}\ {\Big
{)}}\ {\bar {x}}\ \ \ {\to}\ \ \ dI\ = 0$\ ,}
\end{center}
then the {\bf Schr\"{o}dinger equation continuos measurement}
{\underline {has}} an {\bf Ermakov-Lewis invariant} for the
$TDHO$.
\end{document}